# Search for the Production of Single Sleptons through $R$-parity Violation in $p\bar{p}$ Collisions at $\sqrt{s}$ =1.8 TeV

V.M. Abazov,[22] B. Abbott,[56] A. Abdesselam,[11] M. Abolins,[49] V. Abramov,[25] B.S. Acharya,[17] D.L. Adams,[54] M. Adams,[36] S.N. Ahmed,[21] G.D. Alexeev,[22] A. Alton,[48] G.A. Alves,[2] E.W. Anderson,[41] Y. Arnoud,[9] C. Avila,[5] V.V. Babintsev,[25] L. Babukhadia,[53] T.C. Bacon,[27] A. Baden,[45] B. Baldin,[35] P.W. Balm,[20] S. Banerjee,[17] E. Barberis,[29] P. Baringer,[42] J. Barreto,[2] J.F. Bartlett,[35] U. Bassler,[12] D. Bauer,[27] A. Bean,[42] F. Beaudette,[11] M. Begel,[52] A. Belyaev,[34] S.B. Beri,[15] G. Bernardi,[12] I. Bertram,[26] A. Besson,[9] R. Beuselinck,[27] V.A. Bezzubov,[25] P.C. Bhat,[35] V. Bhatnagar,[15] M. Bhattacharjee,[53] G. Blazey,[37] F. Blekman,[20] S. Blessing,[34] A. Boehnlein,[35] N.I. Bojko,[25] T.A. Bolton,[43] F. Borcherding,[35] K. Bos,[20] T. Bose,[51] A. Brandt,[58] R. Breedon,[30] G. Briskin,[57] R. Brock,[49] G. Brooijmans,[35] A. Bross,[35] D. Buchholz,[38] M. Buehler,[36] V. Buescher,[14] V.S. Burtovoi,[25] J.M. Butler,[46] F. Canelli,[52] W. Carvalho,[3] D. Casey,[49] Z. Casilum,[53] H. Castilla-Valdez,[19] D. Chakraborty,[37] K.M. Chan,[52] S.V. Chekulaev,[25] D.K. Cho,[52] S. Choi,[33] S. Chopra,[54] J.H. Christenson,[35] D. Claes,[50] A.R. Clark,[29] L. Coney,[40] B. Connolly,[34] W.E. Cooper,[35] D. Coppage,[42] S. Crépé-Renaudin,[9] M.A.C. Cummings,[37] D. Cutts,[57] G.A. Davis,[52] K. De,[58] S.J. de Jong,[21] M. Demarteau,[35] R. Demina,[43] P. Demine,[9] D. Denisov,[35] S.P. Denisov,[25] S. Desai,[53] H.T. Diehl,[35] M. Diesburg,[35] S. Doulas,[47] Y. Ducros,[13] L.V. Dudko,[24] S. Duensing,[21] L. Duflot,[11] S.R. Dugad,[17] A. Duperrin,[10] A. Dyshkant,[37] D. Edmunds,[49] J. Ellison,[33] J.T. Eltzroth,[58] V.D. Elvira,[35] R. Engelmann,[53] S. Eno,[45] G. Eppley,[60] P. Ermolov,[24] O.V. Eroshin,[25] J. Estrada,[52] H. Evans,[51] V.N. Evdokimov,[25] D. Fein,[28] T. Ferbel,[52] F. Filthaut,[21] H.E. Fisk,[35] Y. Fisyak,[54] E. Flattum,[35] F. Fleuret,[12] M. Fortner,[37] H. Fox,[38] S. Fu,[51] S. Fuess,[35] E. Gallas,[35] A.N. Galyaev,[25] M. Gao,[51] V. Gavrilov,[23] R.J. Genik II,[26] K. Genser,[35] C.E. Gerber,[36] Y. Gershtein,[57] R. Gilmartin,[34] G. Ginther,[52] B. Gómez,[5] P.I. Goncharov,[25] H. Gordon,[54] L.T. Goss,[59] K. Gounder,[35] A. Goussiou,[27] N. Graf,[54] P.D. Grannis,[53] J.A. Green,[41] H. Greenlee,[35] Z.D. Greenwood,[44] S. Grinstein,[1] L. Groer,[51] S. Grünendahl,[35] A. Gupta,[17] S.N. Gurzhiev,[25] G. Gutierrez,[35] P. Gutierrez,[56] N.J. Hadley,[45] H. Haggerty,[35] S. Hagopian,[34] V. Hagopian,[34] R.E. Hall,[31] S. Hansen,[35] J.M. Hauptman,[41] C. Hays,[51] C. Hebert,[42] D. Hedin,[37] J.M. Heinmiller,[36] A.P. Heinson,[33] U. Heintz,[46] M.D. Hildreth,[40] R. Hirosky,[61] J.D. Hobbs,[53] B. Hoeneisen,[8] Y. Huang,[48] I. Iashvili,[33] R. Illingworth,[27] A.S. Ito,[35] M. Jaffré,[11] S. Jain,[17] R. Jesik,[27] K. Johns,[28] M. Johnson,[35] A. Jonckheere,[35] H. Jöstlein,[35] A. Juste,[35] W. Kahl,[43] S. Kahn,[54] E. Kajfasz,[10] A.M. Kalinin,[22] D. Karmanov,[24] D. Karmgard,[40] R. Kehoe,[49] A. Khanov,[43] A. Kharchilava,[40] S.K. Kim,[18] B. Klima,[35] B. Knuteson,[29] W. Ko,[30] J.M. Kohli,[15] A.V. Kostritskiy,[25] J. Kotcher,[54] B. Kothari,[51] A.V. Kozelov,[25] E.A. Kozlovsky,[25] J. Krane,[41] M.R. Krishnaswamy,[17] P. Krivkova,[6] S. Krzywdzinski,[35] M. Kubantsev,[43] S. Kuleshov,[23] Y. Kulik,[35] S. Kunori,[45] A. Kupco,[7] V.E. Kuznetsov,[33] G. Landsberg,[57] W.M. Lee,[34] A. Leflat,[24] C. Leggett,[29] F. Lehner,[35,*] C. Leonidopoulos,[51] J. Li,[58] Q.Z. Li,[35] J.G.R. Lima,[3] D. Lincoln,[35] S.L. Linn,[34] J. Linnemann,[49] R. Lipton,[35] A. Lucotte,[9] L. Lueking,[35] C. Lundstedt,[50] C. Luo,[39] A.K.A. Maciel,[37] R.J. Madaras,[29] V.L. Malyshev,[22] V. Manankov,[24] H.S. Mao,[4] T. Marshall,[39] M.I. Martin,[37] A.A. Mayorov,[25] R. McCarthy,[53] T. McMahon,[55] H.L. Melanson,[35] M. Merkin,[24] K.W. Merritt,[35] C. Miao,[57] H. Miettinen,[60] D. Mihalcea,[37] C.S. Mishra,[35] N. Mokhov,[35] N.K. Mondal,[17] H.E. Montgomery,[35] R.W. Moore,[49] M. Mostafa,[1] H. da Motta,[2] Y.D. Mutaf,[53] E. Nagy,[10] F. Nang,[28] M. Narain,[46] V.S. Narasimham,[17] N.A. Naumann,[21] H.A. Neal,[48] J.P. Negret,[5] A. Nomerotski,[35] T. Nunnemann,[35] D. O'Neil,[49] V. Oguri,[3] B. Olivier,[12] N. Oshima,[35] P. Padley,[60] K. Papageorgiou,[36] N. Parashar,[47] R. Partridge,[57] N. Parua,[53] A. Patwa,[53] O. Peters,[20] P. Pétroff,[11] R. Piegaia,[1] B.G. Pope,[49] E. Popkov,[46] H.B. Prosper,[34] S. Protopopescu,[54] M.B. Przybycien,[38,†] J. Qian,[48] R. Raja,[35] S. Rajagopalan,[54] P.A. Rapidis,[35] N.W. Reay,[43] S. Reucroft,[47] M. Ridel,[11] M. Rijssenbeek,[53] F. Rizatdinova,[43] T. Rockwell,[49] M. Roco,[35] C. Royon,[13] P. Rubinov,[35] R. Ruchti,[40] J. Rutherfoord,[28] B.M. Sabirov,[22] G. Sajot,[9] A. Santoro,[3] L. Sawyer,[44] R.D. Schamberger,[53] H. Schellman,[38] A. Schwartzman,[1] E. Shabalina,[36] R.K. Shivpuri,[16] D. Shpakov,[47] M. Shupe,[28] R.A. Sidwell,[43] V. Simak,[7] H. Singh,[33] V. Sirotenko,[35] P. Slattery,[52] R.P. Smith,[35] R. Snihur,[38] G.R. Snow,[50] J. Snow,[55] S. Snyder,[54] J. Solomon,[36] Y. Song,[58] V. Sorín,[1] M. Sosebee,[58] N. Sotnikova,[24] K. Soustruznik,[6] M. Souza,[2] N.R. Stanton,[43] G. Steinbrück,[51] R.W. Stephens,[58] D. Stoker,[32] V. Stolin,[23] A. Stone,[44] D.A. Stoyanova,[25] M.A. Strang,[58] M. Strauss,[56] M. Strovink,[29] L. Stutte,[35] A. Sznajder,[3] M. Talby,[10] W. Taylor,[53] S. Tentindo-Repond,[34] S.M. Tripathi,[30] T.G. Trippe,[29] A.S. Turcot,[54] P.M. Tuts,[51] V. Vaniev,[25] R. Van Kooten,[39] N. Varelas,[36] L.S. Vertogradov,[22] F. Villeneuve-Seguier,[10] A.A. Volkov,[25] A.P. Vorobiev,[25] H.D. Wahl,[34] H. Wang,[38] Z.-M. Wang,[53] J. Warchol,[40] G. Watts,[62] M. Wayne,[40] H. Weerts,[49] A. White,[58] J.T. White,[59] D. Whiteson,[29] D.A. Wijngaarden,[21] S. Willis,[37] S.J. Wimpenny,[33] J. Womersley,[35] D.R. Wood,[47] Q. Xu,[48] R. Yamada,[35] P. Yamin,[54] T. Yasuda,[35] Y.A. Yatsunenko,[22] K. Yip,[54] S. Youssef,[34] J. Yu,[58] M. Zanabria,[5] X. Zhang,[56] H. Zheng,[40] B. Zhou,[48] Z. Zhou,[41] M. Zielinski,[52] D. Zieminska,[39] A. Zieminski,[39] V. Zutshi,[37] E.G. Zverev,[24] and A. Zylberstejn[13]

(DØ Collaboration)

[1] *Universidad de Buenos Aires, Buenos Aires, Argentina*
[2] *LAFEX, Centro Brasileiro de Pesquisas Físicas, Rio de Janeiro, Brazil*
[3] *Universidade do Estado do Rio de Janeiro, Rio de Janeiro, Brazil*
[4] *Institute of High Energy Physics, Beijing, People's Republic of China*
[5] *Universidad de los Andes, Bogotá, Colombia*
[6] *Charles University, Center for Particle Physics, Prague, Czech Republic*
[7] *Institute of Physics, Academy of Sciences, Center for Particle Physics, Prague, Czech Republic*
[8] *Universidad San Francisco de Quito, Quito, Ecuador*
[9] *Institut des Sciences Nucléaires, IN2P3-CNRS, Universite de Grenoble 1, Grenoble, France*
[10] *CPPM, IN2P3-CNRS, Université de la Méditerranée, Marseille, France*
[11] *Laboratoire de l'Accélérateur Linéaire, IN2P3-CNRS, Orsay, France*
[12] *LPNHE, Universités Paris VI and VII, IN2P3-CNRS, Paris, France*
[13] *DAPNIA/Service de Physique des Particules, CEA, Saclay, France*
[14] *Universität Mainz, Institut für Physik, Mainz, Germany*
[15] *Panjab University, Chandigarh, India*
[16] *Delhi University, Delhi, India*
[17] *Tata Institute of Fundamental Research, Mumbai, India*
[18] *Seoul National University, Seoul, Korea*
[19] *CINVESTAV, Mexico City, Mexico*
[20] *FOM-Institute NIKHEF and University of Amsterdam/NIKHEF, Amsterdam, The Netherlands*
[21] *University of Nijmegen/NIKHEF, Nijmegen, The Netherlands*
[22] *Joint Institute for Nuclear Research, Dubna, Russia*
[23] *Institute for Theoretical and Experimental Physics, Moscow, Russia*
[24] *Moscow State University, Moscow, Russia*
[25] *Institute for High Energy Physics, Protvino, Russia*
[26] *Lancaster University, Lancaster, United Kingdom*
[27] *Imperial College, London, United Kingdom*
[28] *University of Arizona, Tucson, Arizona 85721*
[29] *Lawrence Berkeley National Laboratory and University of California, Berkeley, California 94720*
[30] *University of California, Davis, California 95616*
[31] *California State University, Fresno, California 93740*
[32] *University of California, Irvine, California 92697*
[33] *University of California, Riverside, California 92521*
[34] *Florida State University, Tallahassee, Florida 32306*
[35] *Fermi National Accelerator Laboratory, Batavia, Illinois 60510*
[36] *University of Illinois at Chicago, Chicago, Illinois 60607*
[37] *Northern Illinois University, DeKalb, Illinois 60115*
[38] *Northwestern University, Evanston, Illinois 60208*
[39] *Indiana University, Bloomington, Indiana 47405*
[40] *University of Notre Dame, Notre Dame, Indiana 46556*
[41] *Iowa State University, Ames, Iowa 50011*
[42] *University of Kansas, Lawrence, Kansas 66045*
[43] *Kansas State University, Manhattan, Kansas 66506*
[44] *Louisiana Tech University, Ruston, Louisiana 71272*
[45] *University of Maryland, College Park, Maryland 20742*
[46] *Boston University, Boston, Massachusetts 02215*
[47] *Northeastern University, Boston, Massachusetts 02115*
[48] *University of Michigan, Ann Arbor, Michigan 48109*
[49] *Michigan State University, East Lansing, Michigan 48824*
[50] *University of Nebraska, Lincoln, Nebraska 68588*
[51] *Columbia University, New York, New York 10027*
[52] *University of Rochester, Rochester, New York 14627*
[53] *State University of New York, Stony Brook, New York 11794*
[54] *Brookhaven National Laboratory, Upton, New York 11973*
[55] *Langston University, Langston, Oklahoma 73050*
[56] *University of Oklahoma, Norman, Oklahoma 73019*
[57] *Brown University, Providence, Rhode Island 02912*
[58] *University of Texas, Arlington, Texas 76019*
[59] *Texas A&M University, College Station, Texas 77843*

[60]*Rice University, Houston, Texas 77005*
[61]*University of Virginia, Charlottesville, Virginia 22901*
[62]*University of Washington, Seattle, Washington 98195*

We report the first search for supersymmetric particles via $s$-channel production and decay of smuons or muon sneutrinos at hadronic colliders. The data for the two-muon and two-jets final states were collected by the DØ experiment, and correspond to an integrated luminosity of 94±5 $\mathrm{pb}^{-1}$. Assuming that $R$-parity is violated via the single coupling $\lambda'_{211}$, the number of candidate events is in agreement with expectation from the standard model. Exclusion contours are given in the $(m_0, m_{1/2})$ and $(m_{\tilde{\chi}}, m_{\tilde{\nu}})$ planes for $\lambda'_{211}$=0.09, 0.08 and 0.07.

Events with at least two muons and two hadronic jets in $p\bar{p}$ collisions provide a good sample in which to search for new physics because the contribution from standard-model processes to such states is rather small. Any excess in such topologies can be attributed to a signal from $R$-parity violating supersymmetry (SUSY), where $R$-parity is not conserved either in the production or in the decay of sparticles.

$R$-parity of any particle [1] is defined as $R_p = (-1)^{3B+L+2S}$, where $B$, $L$ and $S$ are the baryon, lepton and spin quantum numbers. $R_p$ equals $+1$ for SM particles and $-1$ for supersymmetric partners. The conservation of $R$-parity is often assumed in experimental searches, because, without that, simultaneous lepton and baryon number violation would lead to rapid proton decay. However, this argument can be circumvented if lepton and baryon number conservation are treated independently.

In supersymmetry, $R$-parity violation ($\not{R}_p$) can occur through terms in the superpotential, that are trilinear in quark and lepton superfields [1]:

$$\lambda_{ijk} L_i L_j \bar{E}^c_k + \lambda'_{ijk} L_i Q_j \bar{D}^c_k + \lambda''_{ijk} \bar{U}^c_i \bar{D}^c_j \bar{D}^c_k, \qquad (1)$$

where $i, j$, and $k$ are family indices; $L$ and $Q$ are the SU(2)-doublet lepton and quark superfields; $E$, $U$, and $D$ are the singlet-lepton, up-quark, and down-quark superfields, respectively.

Such $\not{R}_p$ couplings offer the possibility of producing single supersymmetric particles [2], which is not the case for $R_p$-conserving supersymmetric models, in which particles and sparticles are always produced in pairs. Although the $\not{R}_p$ coupling constants are severely constrained by low-energy experimental bounds [3,4], $s$-channel production and decay of sparticles can have a substantial cross section at lepton and hadron colliders [5,6].

At $p\bar{p}$ colliders, either a sneutrino ($\tilde{\nu}$) or a charged slepton ($\tilde{l}$) can be produced in the $s$-channel via $\lambda'_{ijk}$ coupling. In most SUSY models, the slepton has two possible $R_p$-conserving gauge decays: either into a chargino $\tilde{\chi}^{\pm}$ or a neutralino $\tilde{\chi}^0$. These are favored over $\not{R}_p$ decay because of the small value of the coupling for the latter [5]. Consequently, for a single dominant $\lambda'_{ijk}$ coupling, production of a slepton (smuon or muon sneutrino) provides either a chargino or a neutralino, together with either a charged lepton or a neutrino, in the final state.

In this Letter, we consider the resonant production of a muon sneutrino or a smuon via $\lambda'_{211}$ coupling which leads to a final state containing at least two muons and two jets. From low-energy measurements the $\lambda'_{211}$ coupling is constrained to be less than $0.06/100(\mathrm{GeV/c^2}) \times m_{\tilde{d_R}}$ [7], where $m_{\tilde{d_R}}$ is the mass of the $\tilde{d_R}$-squark. The lightest supersymmetric particle (LSP) is assumed to be the lightest neutralino. We also assume that all sparticles cascade-decay into the neutralino, which decays through the dominant $\not{R}_p$ $\lambda'_{211}$ coupling. Hence, ultimately, all SUSY particles decay into the lightest neutralino, which decays into two jets and a muon. The decay of the muon sneutrino into a muon and a chargino, and of the smuon into a muon and a neutralino, therefore lead to at least two muons and two jets in the final state. The decay of the smuon into a neutrino and a chargino can also lead to the same topology, but only when the chargino decays into muon$+X$, and for this reason the contribution of that channel is small (less than 5% of the signal) and neglected in our analysis. The decay of the sneutrino into a neutrino and a neutralino yields only one muon in the final state.

Our framework is the so-called minimal supergravity model (mSUGRA), which assumes the existence of a grand unified gauge theory and family-universal boundary conditions on the supersymmetry breaking parameters. We choose the following five parameters that completely define the model: $m_0$, the universal scalar mass at the unification scale $M_X$; $m_{1/2}$, the universal gaugino mass at $M_X$; $A = A_t = A_b = A_\tau$, the trilinear Yukawa coupling at $M_X$, $sign(\mu)$, the sign of the Higgsino mixing parameter; $\tan\beta = < H_u > / < H_d >$ where $< H_u >$ and $< H_d >$ denote the vacuum expectation values of the two Higgs fields. The dependence of the cross section on different SUSY parameters can be found in Ref. [5].

The data for this analysis were collected during the 1994-1995 Fermilab Tevatron running, at a center-of-mass energy of 1.8 TeV, and correspond to an integrated luminosity of $94 \pm 5$ pb$^{-1}$. The DØ detector is described elsewhere [9]. Here, we outline the performance of the components relevant to this analysis. Jets are identified using the energy deposited in the calorimeter, and reconstructed with a cone algorithm in pseudorapidity ($\eta$) and azimuthal angle ($\phi$) using a radius of 0.5. The calorimeter covers the region of $|\eta| < 4.0$, and provides a resolution for electrons and single hadrons ($\sigma(E)/E$) of $15\%/\sqrt{E}$ and $50\%/\sqrt{E}$, respectively. Muons are detected using both tracking chambers (three layers of proportional drift tubes ($|\eta| < 1.7$), one in front of, and two behind magnetized iron toroids) and through ionization deposited in the calorimeter. The muon momentum resolution is $\sigma(1/p) = 0.18(p-2)/p^2 + 0.003$ (with $p$ in GeV/$c$).

Events are required to satisfy a $\mu$ + jet or $\mu\mu$ + jet trigger. The trigger efficiency is 71% and 50% for central and forward muons, respectively. Muons are required to have a transverse momentum greater than 8 GeV/$c$, and jets are required to have transverse energy exceeding 15 GeV. We apply additional criteria to select two isolated muons and to eliminate cosmic-ray muons. If there are more than two isolated muons (which happens only rarely), only the two leading muons are used in the ensuing analysis.

The signal topologies were generated with the SUSYGEN Monte Carlo program [8] using the cross sections computed in Refs. [5,6] for a wide range of ($m_0$, $m_{1/2}$) masses. For illustration purposes, we choose a reference point in the mSUGRA parameter space: $m_0$=200 GeV/c$^2$, $m_{1/2}$=243 GeV/c$^2$, $\tan\beta = 2$, $A = 0$, and a negative sign for $\mu$. These parameters predict the following sparticle masses: $m_{\tilde{\nu}}$ =263 GeV/c$^2$ , $m_{\tilde{\mu}}$ =269 GeV/c$^2$ , $m_{\tilde{\chi}_1^\pm}$ =207 GeV/c$^2$, and $m_{\tilde{\chi}_1^0}$ =102 GeV/c$^2$. For $\lambda' = 0.09$, the production cross sections are 1.22 pb and 3.34 pb for $\tilde{\nu}$ and $\tilde{\mu}$ production, respectively.

The dominant backgrounds are from $t\bar{t}$, $WW$+jets and $Z$+2 jets events. The $t\bar{t}$ background was generated using PYTHIA [10], with a cross section of 5.9 ± 1.7 pb [11], the $Z$+2 jets background with VECBOS, [12] interfaced with the ISAJET fragmentation code [13], and a cross section of 9.7 ± 0.9 pb. The $WW$+jets background was generated using PYTHIA [10]; it provides a much smaller background than the $t\bar{t}$ and $Z$+2 jets channels. The simulation of the detector was performed using both a full and a parameterized simulation.

We use a neural network to discriminate signal from background in our analysis [14] and we cross-check this with a more standard sequential analysis at several points of the SUSY parameter space. The following quantities are used as inputs to the neural network: the scalar sum of the transverse energies of the two leading jets, the scalar sum of the transverse momenta of the two leading muons, the distance in ($\eta$, $\phi$) space between the two muons, the dimuon mass, the ($\eta$, $\phi$) distance between the most energetic muon and its nearest jet, the aplanarity and the sphericity of the two leading muons and two leading jets in the laboratory frame [15].

The output of the neural network is obtained separately for the sneutrino and the smuon channels. The signal-over-background ratio for the neural network is optimal for an output cutoff of 0.0 for the $\tilde{\nu}$ and $-0.10$ for the $\tilde{\mu}$ analysis.

For the reference point, $6.42 \pm 0.06$ $\tilde{\nu}$ and $\tilde{\mu}$ events are expected. The estimated background of 1.01±0.02 events is consistent with the two events observed in data. The details of the background estimate are given in Table I, with the quoted uncertainties being only statistical.

The systematic errors are shown in Table II. The uncertainties due to jet energy scale and the measurement of the muon $p_T$ are deduced by varying the jet $E_T$ and muon $p_T$ by one standard deviation. We use a fast version of the detector simulation for most of the SUSY points, and the systematic error associated with this procedure is also given in Table II. The last three lines give the final results for the number of events, the overall statistical error, and the overall systematic error. Using a Bayesian method to calculate the level of exclusion [16], our specific reference point is rejected at the 97.7% C.L. for $\lambda'_{211} = 0.09$.

| | |
|---|---|
| $\tilde{\mu}$ | $3.93 \pm 0.05$ |
| $\tilde{\nu}_\mu$ | $2.49 \pm 0.04$ |
| Total expected signal | $6.42 \pm 0.06$ |
| $t\bar{t}$ | $0.27 \pm 0.01$ |
| $Z$+2jets | $0.73 \pm 0.02$ |
| $WW$ + jets | $0.01 \pm 0.00$ |
| Total background | $1.01 \pm 0.02$ |
| Data | 2 |
| CL | 97.7% |

TABLE I. Number of events (expected) for the reference point for signal at $\lambda' = 0.09$, for the background, and the number observed in the data after making all selections.

| Source | Signal | $t\bar{t}$ | $Z$+2 jets |
|---|---|---|---|
| Jet energy scale | 2% | 4% | 5% |
| High $p_T^\mu$ efficiency | 1% | 7% | 4% |
| Cross section | 10% | 30% | 10% |
| Trigger simul. | 5% | 5% | 5% |
| Luminosity | 5% | 5% | 5% |
| fast / full simul. | 1% | 1% | 1% |
| Total events | 6.42 | 0.27 | 0.73 |
| Overall statistics | ± 0.06 | ± 0.01 | ± 0.02 |
| Overall systematics | ± 0.80 | ± 0.09 | ± 0.10 |

TABLE II. Systematic uncertainties on signal and background, and the number of expected events, with their statistical and systematic errors.

To set exclusion contours, we scan the ($m_0$, $m_{1/2}$) plane for three values of the coupling constant $\lambda'_{211}$ = 0.09, 0.08, 0.07, two values of $\tan\beta$ = 2, 5, all for sign($\mu$) = −1. For $\lambda' \geq 0.09$, the coupling $100\lambda'/m_{\tilde{d}_R}$ is almost completely excluded by earlier experiments [7] in our domain of sensitivity in $m_{\tilde{d}_R}$. The resulting exclusion contours at the 95% C.L. are shown in Figs. 1 and 2 in the ($m_0$, $m_{1/2}$) plane. The most interesting feature, is the exclusion of $m_{1/2}$ values up to 260 GeV/c$^2$ for $\tan\beta$=2 and $\lambda'_{211} = 0.09$, and the exclusion of $\tilde{\nu}$ and $\tilde{\mu}$ with masses up to 280 GeV/c$^2$.

For low values of $m_0$ and $m_{1/2}$, the smuon mass is close to the chargino or neutralino mass, the $p_T$ spectrum of the muons is soft, and the search is inefficient. For $\mu$ >0 and higher values of $\tan\beta$, the sensitivity of our reach is expected to decrease due to the fact that the photino component of the LSP becomes small, resulting in the decrease of the branching fraction of the LSP into muons. In addition, charginos and neutralinos become light, resulting in events with softer muons and jets that fail the kinematic requirements.

To conclude, a search for single smuon and single muon sneutrino production in the mSUGRA model with R-parity violation, has been performed for the first time at the Tevatron. We exclude $m_{1/2}$ values up to 260 GeV (the excluded value of $m_{1/2}$ depends on the value of $m_0$) and sneutrino and smuon masses up to 280 GeV. The excluded domain in the ($m_0$, $m_{1/2}$) plane extends the region excluded using the dielectron channel [17].

We thank G. Moreau, M. Chemtob, R. Peschanski and C. Savoy for useful discussions. We thank the staffs at Fermilab and collaborating institutions, and acknowledge support from the Department of Energy and National Science Foundation (USA), Commissariat à L'Energie Atomique and CNRS/Institut National de Physique Nucléaire et de Physique des Particules (France), Ministry for Science and Technology and Ministry for Atomic Energy (Russia), CAPES and CNPq (Brazil), Departments of Atomic Energy and Science and Education (India), Colciencias (Colombia), CONACyT (Mexico), Ministry of Education and KOSEF (Korea), CONICET and UBACyT (Argentina), The Foundation for Fundamental Research on Matter (The Netherlands), PPARC (United Kingdom), Ministry of Education (Czech Republic), A.P. Sloan Foundation, and the Research Corporation.

---

* Visitor from University of Zurich, Zurich, Switzerland.
† Visitor from Institute of Nuclear Physics, Krakow, Poland.

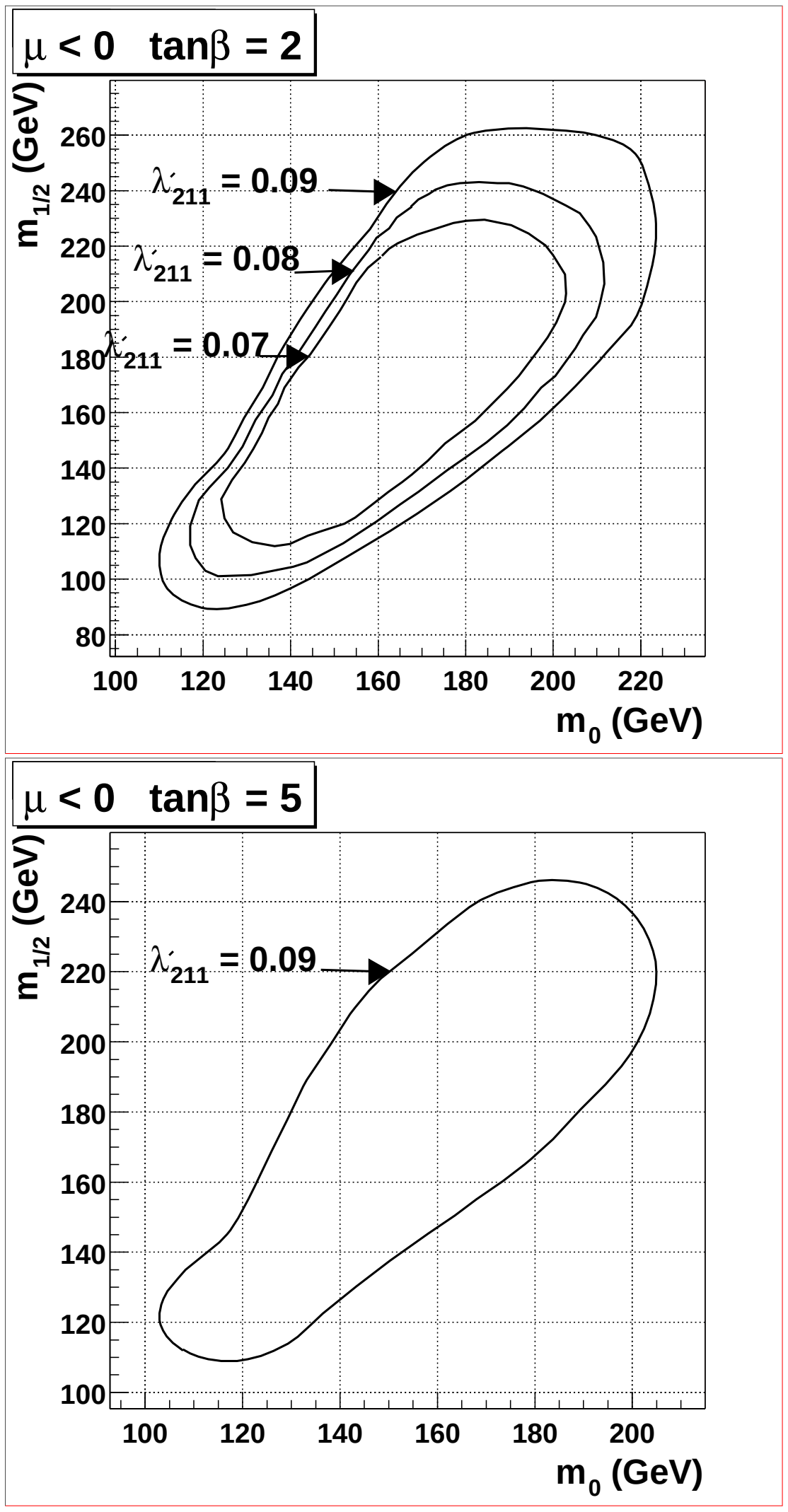

FIG. 1. Exclusion contours at the 95% C.L. in the ($m_0$, $m_{1/2}$) plane. The top figure shows the exclusion contours for $\tan\beta = 2$, $\lambda'_{211}$ =0.09, 0.08 and 0.07. The bottom figure shows the exclusion contour for $\tan\beta = 5$, but only for $\lambda'_{211} = 0.09$, because the smaller couplings do not provide a region of 95% C.L. exclusion.

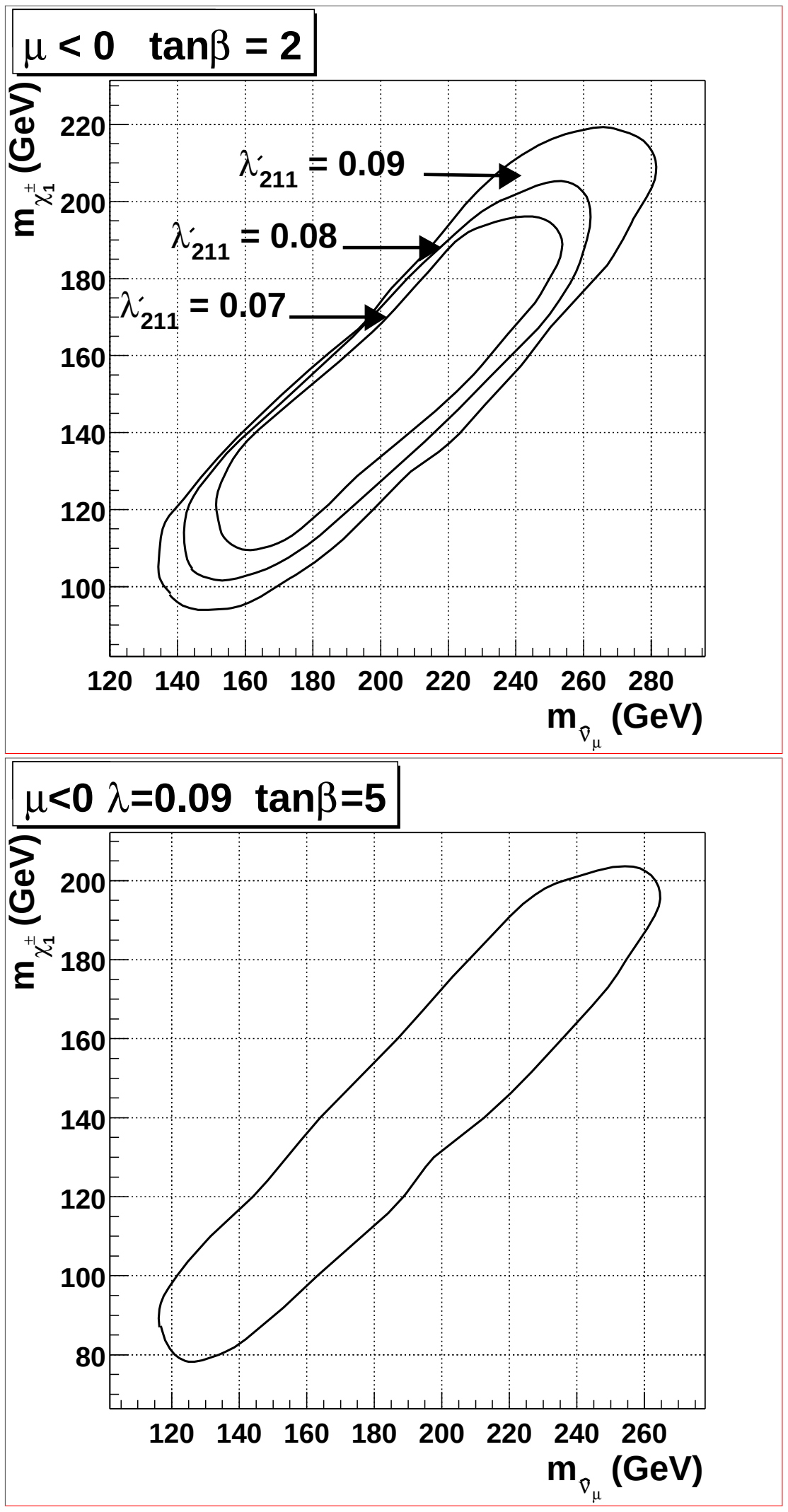

FIG. 2. Exclusion contours at the 95% C.L. in the ($m_{\tilde{\mu}}/m_{\tilde{\nu}}$, $m_{\chi^+}$) plane. The top figure is for $\tan\beta = 2$, and three values of $\lambda'_{211}$, while the bottom figure is for $\tan\beta = 5$ and $\lambda'_{211} = 0.09$. We give all contour plots as a function of the sneutrino mass. Because, for any given set of parameters, the sneutrino mass is very close to the smuon mass, the smuon contour plots lie very close to the sneutrino results, and are therefore not shown.